\documentclass[11pt]{llncs}
\usepackage[margin=1in]{geometry}
\usepackage{times}  
\usepackage{helvet}  
\usepackage{courier}  
\usepackage[hyphens]{url}  
\usepackage{graphicx} 
\usepackage{natbib}

\usepackage{caption}
\usepackage{algorithm}
\usepackage{algorithmic}
\usepackage{newfloat}
\usepackage{listings}
\DeclareCaptionStyle{ruled}{labelfont=normalfont,labelsep=colon,strut=off} 
\floatstyle{ruled}
\newfloat{listing}{tb}{lst}{}
\floatname{listing}{Listing}
\title{Facility Location for Congesting Commuters \\
and Generalizing the Cost-Distance Problem%
}

\author{Thanasis Lianeas$^{1}$ \and
Marios Mertzanidis$^{2}$ \and
Aikaterini Nikolidaki$^{3}$}
\authorrunning{Thanasis Lianeas,
Marios Mertzanidis,
Aikaterini Nikolidaki}

\institute{$^1$University of West Attica, $^2$Purdue University, $^3$National Technical University of Athens.\\
\email{lianeas@corelab.ntua.gr, mmertzan@purdue.edu, aiknikol@mail.ntua.gr}}

\usepackage{graphicx}
\usepackage{amsmath}
\usepackage{mathtools}
\usepackage{amssymb}
\usepackage{subfig}

\usepackage{xspace}
\newcommand{\flc}{\textit{FLCC}\xspace}
\newcommand{\fls}{\textit{FLSC}\xspace}
\newcommand{\icd}{I_{cd}\xspace}
\newcommand{\paths}{\mathcal{P}}
\newcommand{\reals}{\mathbb{R}}

\usepackage{xcolor}

\begin{document}

\maketitle

\begin{abstract}
In Facility Location problems there are agents that should be connected to facilities and  locations where facilities may be opened so that agents can connect to them.
We depart from Uncapacitated Facility Location and by assuming that the connection costs of agents to facilities are congestion dependent, we define a novel problem, namely, Facility Location for Congesting (Selfish) Commuters. The connection costs of agents to facilities come as a result of how the agents commute to reach the facilities in an underlying network with cost  functions on the edges. 
Inapproximability results follow from the related literature and thus approximate solutions is all we can hope for. For when the cost functions are nondecreasing we employ in a novel way an approximate version of Caratheodory's Theorem \cite{ApproxCaratheodoryBarman} to show how approximate solutions for different versions of the problem can be derived.
For when the cost functions are nonincreasing  we show how this problem generalizes the Cost-Distance problem \cite{Bicriteria} and provide an algorithm that for this more general case achieves the same approximation guarantees.
\end{abstract}



\section{Introduction}
Facility Location problems are among the well studied problems in Theoretical  Computer Science (see, e.g., the related work section) and other fields \cite{FLbook1_Man_Science,FLbook2_ThAndApp}. In such problems, there are some facilities to be opened in available locations and agents need to connect to them. Opening facilities usually comes with a facility cost and connecting agents to their closest facility comes with connection costs. The goal is to minimize the sum of these costs. The connection costs are usually fixed and ignore congestion effects when connecting to the facilities. In this work, we assume that there are congestion effects  when  the agents commute to connect to the facilities.

Congestion Games are one of the most studied classes of games in Algorithmic Game Theory. In these games, agents choose subsets of resources and incur  costs based on the congestion on the resources they have chosen. In Network Congestion Games, the very basic version of Congestion Games, there is an underlying network and agents choose paths in the network that connect their sources to their targets. Agents are selfish, aiming to minimize their costs without caring for any type of social welfare, which could be the goal of a central organizer. For that reason, it makes sense to examine both outcomes where agents are satisfied by being on shortest paths, thus being at equilibrium, and outcomes where agents are assigned so that the social welfare, in any chosen sense,  is minimized.

Departing from Uncapacitated Facility Location we define Facility Location for Congesting/Selfish Commuters. We assume there is an underlying network on the nodes of which facilities may be opened with a node-specific cost. Some nodes of the network are source nodes from where infinitesimally small agents, that form demands, depart to reach facilities.  For any given set of facilities, there are many flows that may route the demands to the facilities, with the congestion cost varying among them. The goal is to minimize the sum of the facilities opening costs for the opened facilities plus the routing costs of the demands. When minimizing, similar to Congestion Games, we consider two cases for the flows, asking either to be simply feasible  or to be equilibrium flows. In the first case, we say that we deal with Facility Location for Congesting Commuters and in the second case we say that we deal with Facility Location for Selfish Commuters.

The congestion costs are modeled using cost functions that come with the edges.  Both cases of nondecreasing and nonincreasing cost functions on the edges have been considered in the Congestion Games literature. Interestingly, Facility Location for Congesting Commuters with nonincreasing cost functions generalizes the Cost-Distance problem \cite{Bicriteria}. In this problem, there is an underlying undirected network where demands are to be routed  to some target and every edge comes with two types of costs. The first type of cost can be seen as a cost for building the edge and the second as a cost for traversing the edge. The goal is to return among all Steiner trees that connect the sources to the target the one that minimizes the building plus the routing costs.

\smallskip

\noindent \textbf{Contribution:}
In this work we define Facility Location for Congesting Commuters (\flc in short), a problem generalizing Facility Location Problems by assuming congestion effects that affect the agents when traveling/commuting to connect to their facilities. Moving one step further, we also define Facility Location for Selfish Commuters (\fls in short)  where the agents are not centrally controlled and are not assigned to paths in any convenient way. Instead,  they are selfish wanting to travel in shortest paths, thus we assume that they ought to be at equilibrium. If we see these problems through an Algorithmic Game Theory lens, it is like having Network Congestion Games where the targets for the demands have to be specified (with some cost) affecting any optimization task we wish to solve.

One can think of many variants for  both \flc and \fls since the underlying network  may
be directed or undirected, may be a single-commodity, a single-source multi-commodity or a multi-commodity network and may have  nonincreasing or nondecreasing cost functions, while, in addition, the opening costs for the facilities may be left arbitrary or be constrained, e.g., we may require to have the same  opening cost for all facilities. 
We note that when the cost functions are constants the two problems coincide and reduce to Uncapacitated Facility Location for which hardness of approximation results are known \cite{guha_khuller_1999}.  Thus, we can aim only for approximation algorithms for these problems. One can show (Appendix \ref{app:FLtechniquesFail}) that the techniques used to tackle Facility Location hardness results (Local search and Linear Programming) fail in this more general setting (at least if applied in a straightforward way). 

Our first result concerns directed, single-source multi-commodity networks with nondecreasing  latency functions and arbitrary opening costs. We employ ideas used in Congestion Games that work when the latency functions are $a$-Lipschitz, appplying an approximate version of Caratheodory's theorem (Theorem \ref{theorem:ApproxCaratheodoryAPP}) to provide an algorithm (Section \ref{sec:carath}) that performs well whenever the maximum length among the paths is $o(|V|)$  
(Theorem \ref{thm:CarAppliedForIncreAPP}). The presented algorithm approximately solves \fls, but a similar and simpler approach can be used to approximately solve \flc.  We note that sparsification techniques similar to ours  have been used in the past \cite{AlthoferFotakis,FotLianDimSer_caraththoedo} yet this is the first time that such techniques are applied for multi-target instances,  where additionally the optimal placement for the targets, i.e., the facilities, has to be determined as well.

Our second result concerns undirected multi-commodity networks with nonincreasing cost functions and common opening costs for the facilities.  In this case, as discussed already and as we show in Theorem \ref{theorem:1.4}, \flc generalizes the Cost-Distance problem \cite{Bicriteria} for which  Chuzhoy et al.\cite{chuzhoy_gupta_noar_sinha_2008} have shown an $\Omega(\log \log|S|)$ hard to approximate result, where $S$ is the set of sources. Thus, this hardness result also holds for our more general setting. We propose an approximation algorithm (Subsection \ref{sec:algoDec}) for \flc that is influenced by the algorithm of   Meyerson et al. \cite{Bicriteria} but, in order to make the algorithm work for multicommodity networks and more general latency functions, it uses a novel matching mechanism (for matchings that have to be done therein). This complicates the analysis but is sufficient to provide the same approximation guarantees (Theorem \ref{theorem:logS}). 
The proposed algorithm works for \flc but can also be used for \fls with the factor of approximation multiplied by $PoA(D)$, i.e., the Price of Anarchy \cite{PoA}  for cost functions in class $D$ \cite{DBLP:journals/mor/CorreaSM04}. Conceptually, $PoA(D)$ captures how much worse can an equilibrium flow be when compared to the optimal flow for the cost functions used.

\smallskip
\noindent \textbf{Related Work:} Facility Location problems are among the most well-studied problems in algorithmic literature. Depending on the costs and constraints of the problem different variants of facility location emerge. Our main problem can be seen as a generalization of the Metric Uncapacitated Facility Location problem. The first constant factor approximation algorithm for this problem was provided by  Shmoys et al. \cite{shmoys_tardos_aardal_1997}. They provided a guarantee of 3.16 times the optimal cost. Jain and  Vazirani \cite{JainVazirani} use a Primal-Dual technique to achieve a $3$-approximate algorithm.  Chudak and Shmoys \cite{chudak_shmoys_2003} found a $(1+2/e)$-approximate algorithm and finally Li  \cite{li_2013} provided a 1.488-approximation algorithm for the problem. As far as inapproximability results are concerned, the work of Guha and Khuller \cite{guha_khuller_1999} prove that it is NP-Hard to approximate this problem with a factor better than $1.463$. 

We also examine a k-median variant of our main problem.  Charikar et al. \cite{charikar_chekuri_goel_guha_1998} were able to provide a $O(\log(n) \cdot \log\log(n))$-approximation algorithm for the k-median facility location, which was obtained by derandomizing and refining an algorithm proposed by Bartal \cite{bartal_1996,bartal_1998}.  Charikar et al. \cite{charikar_guha_tardos_shmoys_2002} were able to provide the first constant approximation algorithm for the k-median problem with a guarantee of $6\frac{2}{3}$. Finally, Arya et al. \cite{arya_garg_khandekar_meyerson_munagala_pandit_2004} provided a $3+\epsilon$ approximation algorithm. One can also derive a $(1+\frac{2}{e}-\epsilon)$ inapproximability result by adapting the proof of hardness for the facility location problem provided by Guha and Khuller \cite{guha_khuller_1999} (this was observed by Jain et al.  \cite{jain_mahdian_markakis_saberi_vazirani_2003}).
Additionally many other variants of facility location problems have been studied such as Multi-Level Facility Location \cite{aardal_chudak_shmoys_1999,Hierarchical}, Connected Facility Location \cite{karger_minkoff_2000,gupta_kleinberg_kumar_rastogi_yener_2001,swamy_kumar_2004,Simple;And;Better}, and Universal Facility Location \cite{vygen_2007,angel_thang_regnault_2014,UFL:7.88,Pandit:Disertation,UFL:5}.

Facility Location problems where congestion effects are present have been studied in the past although in those works the congestion does not affect the connection costs, while, additionally, such works are mainly experimental \cite{HajiaghayiMM03,MARIANOV200832,DBLP:journals/ior/DanM19,DBLP:journals/msom/BaronBK08,SeifFacLocCong,EshaghiFacLocCong,ChakrabortFacLocCong,WANG2023109167,JALILIMARAND2024442}. Instead, our work is purely theoretical and to the best of our knowledge is the first to consider that the congestion affects the connection costs, and it does so in a complex way  similar to that assumed in  (network) Congestion Games.

Congestion Games \cite{Rosenthal1973} provide a natural model for non-cooperative
resource allocation in large-scale communication networks and have
been the subject of intensive research in Algorithmic Game Theory. Many variants have been considered in the literature but the one closely related to ours is that of nonatomic network Congestion Games, also known as non atomic Selfish Routing. Through the use of a potential function \cite{Rosenthal1973} equilibrium existence is guaranteed  and its computation  reduces to solving a convex program, which is also the case in our problem. 

%

The  selfish behavior of the users may cause inefficiency, measured using the Price of Anarchy \cite{PoA}. Related to our work, one way to reduce this inefficiency is to find and remove edges that cause the so called Braess Paradox \cite{Bra68,Mur70}, i.e., that removing edges may improve the networks' performance. Braess Paradox seems not an artifact of optimization theory
\cite{Kel08,RouB05,ValiantR10,CY10} and resolving it has been proved hard to approximate even for simple instances \cite{BraessUnsolvable}. This line of work relates to our work in two ways. First, resolving the paradox lies under the umbrella of Network Design, which is also the case for  Facility Location for Congesting/Selfish Commuters. Second, in this work we employ and generalize techniques that have been applied for resolving the paradox \cite{AlthoferFotakis,FotLianDimSer_caraththoedo}.

Facility Location for Congesting/Selfish Commuters generalizes the Cost-Distance Problem. It was introduced by  Meyerson et l. \cite{BicriteriaOLD,Bicriteria}, and generalizes many important problems, e.g., single-sink
buy-at-bulk with variable pipe types between different sets
of nodes, facility location with buy-at-bulk type costs on
edges, constructing single source multicast trees with good
cost and delay properties, and multi-level facility location. In the same work Meyerson et al.  proposed the first known $O(\log |S|)$ randomized approximation algorithm for the single-sink case on a given undirected graph, where $S$ is the set of sources. This algorithm  was subsequently  derandomized by Chekuri et al. \cite{chekuri2001deterministic}.  Chuzoy et al. \cite{chuzhoy_gupta_noar_sinha_2008} presented an $\Omega (\log\log |S|)$ inapproximability result for the single-sink case, which also holds for our problems.


\section{Preliminaries}

For both  \textbf{Facility Location for Congesting Commuters} and \textbf{Facility Location for Selfish Commuters} (\flc and \fls in short, respectively) an instance is defined in a similar way and the difference from one problem to the other is an extra constraint regarding the routing. An instance for these problems consists of a network $G(V,E)$ (directed 
or undirected) 
together with a set of source nodes $S\subseteq V$, a set of continuous non-negative cost functions $\{\ell_e\}_{e\in E}$ for the edges, traffic demands $\{w_i\}_{i\in S}$ for the source nodes, consisting of infinitesimal agents, and a set of costs $\{B_i\}_{i\in V}$ that capture the (possibly different) costs for opening facilities on the nodes. 
%
For convenience we set $|V|=n$ and $|E|=m$ and in the case of single source  instances, i.e., instances where  $S=\{s\}$, w.l.o.g. we assume that $w_s=1$, since cost functions can be adapted appropriately (setting $\ell^{new}_e(x)=\ell_e(w_sx)$).

\smallskip
\noindent \textbf{Cost functions.} 
For every edge $e$, $\ell_e$ maps non-negative reals to non-negative reals, i.e.  $\ell_e:\reals_{\geq0}\rightarrow \reals_{\geq0}$.  We care both for nondecreasing and nonincreasing cost functions. For the first part of this work every $\ell_e$ is assumed to be nondecreasing. For the second part we focus on nonincreasing cost functions, additionally asking for every $\ell_e$ to be such that $x \cdot \ell_e(x)$ is a nondecreasing concave function. This case can be seen as agents sharing the cost of an edge they use, with the total  cost of each edge  increasing if more agents use the edge. In each section it will be clearly noted whether we examine the nondecreasing or nonincreasing variant.

\smallskip
\noindent \textbf{Flows.} The facilities may open on the network's nodes and act as targets to which the traffic demands travel. Once they are opened, the demands travel from their sources to the targets/facilities forming flows. Formally, let $F\subseteq V$  denote the nodes with opened facilities and $\paths_{i,F}$ denote  the network's paths that start from node $i\in S$ and end with a node  in $F$. Also, let $\paths_F=\cup_{i\in S}\paths_{i,F}$. 
Given $F$, a flow $x=\{x_p\}_{p\in \paths_F}$ is a vector assinging nonnegative reals to the paths of $\paths_F$, and it is feasible if for every $i \in S: \sum_{p\in {\paths_{i,F}}} x_p=w_i$. For a ﬂow $x$ and an edge $e$, we let $x_e=\sum_{p:e\in p} x_p$ denote the amount of ﬂow that $x$ routes through $e$. To denote a feasible flow for a set of facilities $F$ we write $x(F)$ and we may write $x$ instead of $x(F)$ whenever $F$ is clear from the context.

\smallskip
\noindent \textbf{Costs}.
Given $F$ and a feasible flow $x$, we say that a path $p$ is used if for all $e\in p:x_e>0$. The cost of a path $p$ under $x$ is the sum of the costs of its edges, i.e., $\ell_p(x)=\sum_{e\in p}\ell_e(x_e)$. The total routing cost of a flow $x$ is $RC(x)=\sum_{p\in \paths} x_p \cdot \ell_p(x) $ or equivalently $RC(x)=\sum_{e\in E} x_e \cdot \ell_e(x_e)$. 

\smallskip
\smallskip
\noindent \textbf{Equilibria}. We say that $x$ is a Nash flow if for all $i\in S$, and for any used path $p\in P_{i,F}$ and any path $p'\in P_{i,F}$ we have $\ell_p(x)\leq \ell_{p'}(x)$, i.e., for every source the corresponding demand is routed through minimum cost paths. We say that $x$ is an $\epsilon$-Nash  flow if for all $i\in S$, and for any used path $p\in P_{i,F}$ and any path $p'\in P_{i,F}$ we have $\ell_p(x)\leq \ell_{p'}(x)+\epsilon$.   Using potential function arguments (for the continuous version of Rosenthal's potential \cite{Rosenthal1973}) one can show that, given $F$, a Nash flow always exists. 


\smallskip
\noindent \textbf{Solutions and Objective.} Given an \flc or an \fls instance, a solution to it is  a set  $F\subseteq V$  determining the nodes on which a facility opens. A solution $F$ is feasible if there exists a feasible flow $x$ (given $F$).  The goal for \flc is to find a feasible  solution that minimizes the sum of the total routing cost, under the opened facilities and searching among all feasible flows, plus the cost for the facilities that we opened,\footnote{We mix routing  and opening costs but we can assume that, e.g., opening costs are multiplied to be measured in routing costs units.}
 i.e., among all feasible $F$ and all feasible flows for $F$ find the one minimizing  
$C(F):=\sum_{e \in E} x_e(F) \cdot \ell_e(x_e(F)) + \sum_{i\in F} B_i.$ For \fls the objective is the same, i.e., minimizing the above total cost, but there is an extra constraint that allows only flows $x(F)$ that are equilibria.

\smallskip
\smallskip
The following definitions serve as reference points for the problems we generalize in this work.

\smallskip
\noindent \textbf{Cost-Distance Problem \cite{Bicriteria}.} 
We are given an undirected graph $G(V,E)$ along with a set of source nodes $S \subseteq V $ which need to be connected to a single sink node $t \in V$.  Each node $s\in S$ comes with an associated demand $w_s$ to be routed to $t$.  Each edge $e\in E$  is equipped with a cost $c_e$  and a length $l_e$. 
The goal is to  find a connected subgraph $G'(V',E')$ of $G$ which contains all the source nodes and the sink and minimizes the sum
$\sum_{e\in E'}c_e+\sum_{s_i\in S} w_s\cdot l(s_i,t)$, where $l(s_i,t)$ denotes the length of the shortest $s_i$-$t$ path in $G'$.

\smallskip
\noindent 
 \textbf{Facility Location  and k-median.}
The problem we depart from is similar to the Metric Uncapacitated Facility Location Problem. We are given an undirected graph with costs $\{c_e\}_{e\in E}$ on the edges. On every vertex $i$ there might be a demand $d_i$ and  a facility can be opened  with opening cost $B_i$. The goal is to open some facilities where the demands can be connected to, so that the sum of the  total facility opening cost  and the weighted cost of connecting the  demands to the facilities is minimized. 
For the k-median  variant there are no opening costs and the goal is to open (up to) $k$ facilities so that the weighted cost of connecting the  demands to the facilities  is minimized.



\section{
Instances with Nondecreasing 
Functions }
\label{sec:carath}
In this section we consider \flc and \fls on directed networks with nondecreasing cost functions. One can show  (see Appendix \ref{app:FLtechniquesFail}) that the approaches that give approximation algorithms for Facility Location problems fail in our more general case (at least if applied in a straightforward way). Instead, inspired by  techniques that have been used in Congestion Games, we restrict our attention on single-source instances, i.e., instances where $|S|=1$. We additionally assume that all cost functions are $a$-Lipschitz. In Appendix \ref{app:CarathMultiFails} we note on why these  techniques are of no use in the multi-source case, i.e., when $|S|>1$.

Our proposed algorithm approximately solves \fls but  can be easily adapted so that it works for \flc. It  makes  use of an approximate version of Caratheodory's theorem to provide an algorithm that solves any such instance and performs well (timewise) whenever 
the maximum length among the paths is $o(|V|)$,
e.g., it runs in polynomial or quasipolynomial  time  when the length of the paths of the network is bounded by a constant or a polylog factor, respectively. Note that this restricts also the number of paths $|P|$.
Since we focus on single-source instances, w.l.o.g. we assume that the total demand is $w_s = 1$.



Caratheodory's theorem  states that if a point $x\in\reals^d$ lies in the convex hull of a set $X=\{x_1,x_2,\ldots,x_{|X|}\}$ then $x$ can be written as the convex combination of (at most) $d+1$ elements of $X$. 
The approximate version of Caratheodory's theorem that we  use is the following.

\begin{theorem}[\cite{ApproxCaratheodoryBarman}]
\label{theorem:ApproxCaratheodoryAPP}
    Let X be a set of vectors $X = \{x_1, ..., x_n\} \subset \mathbb{R}^d$ and $\epsilon > 0$. For every $\mu \in conv(X)$ and $2 \leq p \leq \infty$ there exists an $O(\frac{p \cdot \gamma^2}{\epsilon^2})$-uniform vector $\mu' \in conv(X)$ such that $\| \mu - \mu' \|_p \leq \epsilon$, where $\gamma = \max_{x \in X} \|x\|_p$ and a vector is $k$-uniform when it can be expressed as an average of $k$ vectors of $X$ with replacements allowed, i.e., as $\sum_{j=1}^k\frac{x_{i_1}}{k}$, with $x_{i_1},\ldots,x_{i_k}\in X$.
\end{theorem}

We will apply the theorem for points in  $\reals^{n+m}$ using the $||\cdot||_2$ norm. We consider arbitrary orderings for the $m$ edges and the $n$ nodes of the network and for each path $p \in \paths$ we  create  vector $x_p = (x_{p,1},\ldots, x_{p,m},x_{p,m+1}, \ldots, x_{p,m+n})$ where $x_{p,e} = 1$ if $e \in p$ and $0$ otherwise, for $1\leq e\leq m$, and $x_{p,m+v} = 1$ if $p$ ends in facility $v$ and $0$ otherwise,  for $1\leq v\leq n$. Let $X=\{x_p\}_{p\in\paths}$ be the set of all these created vectors. 

Suppose in the optimal solution we have a set of opened facilities $F^*$ and a corresponding Nash flow $f^*$ that uses a set of paths $\paths'$. Note that, since $w_s=1$, $f^*$ lies in the convex hull of $X'=\{x_p: p \in \paths'\}$ and thus in the convex hull of $X$, i.e., $f^* \in conv(X)$.
Also, assuming a bound $M$ on the length of the paths of $\paths$, $\gamma = \max_{x \in X}\|x\|_2 = \sqrt{1+\max_{p \in P}|p|} \leq \sqrt{M+1}$, as every vector in $X$ has 1's in at most $M$ edge coordinates and one node coordinate.

 Using Theorem \ref{theorem:ApproxCaratheodoryAPP} for set $X$, the $\|\cdot\|_2$ norm and $\epsilon_1>0$ to be defined later, we can see that there exists $k = O(\frac{M}{\epsilon_1^2})$ paths, say  $i_1, i_2, ..., i_k$, such that for flow $\hat{f} = \sum_{j=1}^k \frac{x_{i_j}}{k}$ it holds that $\|f^*-\hat{f}\|_2 \leq \epsilon_1 \Rightarrow |f_e^* - \hat{f_e}| \leq \epsilon_1$. Assuming that the cost functions are $a$-Lipschitz we get $ |\ell_e(f^*) - \ell_e(\hat{f})| \leq a \cdot \epsilon_1 \Rightarrow |\ell_p(f^*) - \ell_p(\hat{f})| \leq a \cdot M \cdot \epsilon_1$, for any path $p\in\paths$. Choosing $\epsilon_1 = \frac{\epsilon}{2 \cdot a \cdot M}$ and considering any path $p$ used by $f^*$ we get $|L(f^*) - \ell_p(\hat{f})| \leq  \frac{\epsilon}{2}$, where $L(f^*)$ is the common path cost at equilibrium $f^*$. This implies that all used  paths' costs are $\frac{\epsilon}{2}$ close to $L(f^*)$ and thus $\epsilon$ close to each other. Additionally, for any unused path $p'$ it holds  $|\ell_{p'}(f^*) - \ell_{p'}(\hat{f})| \leq  \frac{\epsilon}{2}$ and since $\ell_{p'}(f^*)\geq L(f^*) $
we get $ \ell_{p'}(\hat{f})\geq L(f^*)-\frac{\epsilon}{2}$. Thus, $\hat{f}$ is an $\epsilon$-Nash flow.

Putting it all together, flow $\hat{f}$  uses $O(\frac{a^2 \cdot M^3}{\epsilon^2})$ paths, is an $\epsilon$-Nash flow, and for any used path $p\in\paths$: $\ell_p(\hat{f}) \leq L(f^*) + \frac{\epsilon}{2}$. Multiplying both parts by $\hat{f_p}$ (i.e., the amount of flow passing through path $p$ in $\hat{f}$) we get $\hat{f_p} \cdot \ell_p(\hat{f_p}) \leq \hat{f_p} \cdot (L(f^*) + \frac{\epsilon}{2})$. Summing over all $p$ that are being used by $\hat{f}$ we get $\sum_p \hat{f_p} \cdot \ell_p(\hat{f_p}) \leq \sum_p \hat{f_p} \cdot (L(f^*) +\frac{\epsilon}{2}) \Rightarrow RC(\hat{f}) \leq (L(f^*) + \epsilon) \cdot \sum_p \hat{f_p} = L(f^*) + \epsilon = RC(f^*) +\frac{\epsilon}{2}$. 

Last thing to note is that there exists one such $\hat{f}$ that  uses only facilities, say $F'$, that are used in the optimal solution $F^*$. This is because $f^*$ and thus $\hat{f}$ lies in the convex hull of $X'=\{x_p: p \in \paths'\}$, i.e., the used by $f^*$ paths that end up in a node of the network where a facility is opened under $F^*$. Thus, the facilities opening cost in order to  route $\hat{f}$ does not exceed the cost of opening the facilities in $F^*$ implying that in total the routing cost of $\hat{f}$ together with the opening cost of $F'$ is $\leq C(F^*) + \frac{\epsilon}{2}$

Algorithm \ref{algo:Caratheodory_app} exhaustively searches among all flows that may arise as k-combinations of 
paths in $G$. For each of them, it checks which facilities are opened and whether the flow is an $\epsilon$-Nash flow. It returns the $F$ for which the corresponding sum of routing and facilities cost is minimized, and a corresponding $\epsilon$-Nash flow.



\begin{algorithm}[!ht]
\caption{The approximation scheme}  
\label{algo:Caratheodory_app}
  \textbf{Input:} FLSC Instance $G$ and $\alpha, \epsilon >0$\\
  \textbf{Output:} Feasible solution $F$ of $G$ and $ \epsilon$-Nash flow with cost $C\leq C(F^{*}) + \frac{\epsilon}{2}$

  \smallskip\smallskip
  $M$:= the maximum length among the paths of $G$;\\
  $k:=O(\frac{\alpha^{2}M^{3}}{\epsilon^{2}})$;

  \smallskip
  $K$:= the set of all different $k$-combinations of paths of $G$, with replacements allowed;

  \smallskip
  Initialize BestCost:=$\infty$, $F:= $ any set of vertices, and $g:= $any flow;

  \smallskip
\textbf{For} every combination in $K$  \textbf{do}
\begin{itemize}
    \item[--] Let $F'$ be the solution defined by the combination's paths;
    \item[--] Check if the combination's flow $f$, sending $\frac{1}{k}$ units to each path, is an $ \epsilon-$Nash flow for $F'$;
    \item[--] Check if the total routing and facilities cost $C$  is $C\leq BestCost$;
    \item[--] If the answer is yes for both conditions, $BestCost:=C$, $F:=F'$ and $g:=f$;
\end{itemize}
          	        
\textbf{return}\textit{ $F$ and $g$;} \\
\end{algorithm}

\begin{theorem}\label{thm:CarAppliedForIncreAPP}
    For any  \fls instance with $a$-Lipschitz cost functions and any $\epsilon>0$   we can find in time $|\paths|^{O\left(\frac{a^2 \cdot M^3}{\epsilon^2}\right)}\cdot O\big(poly(n)\big)$ 
    a feasible solution $F$ and an $\epsilon$-Nash flow (that routes to these facilities) with  total cost at most $C(F^*) + \frac{\epsilon}{2}$, where $M$ is an upper bound on the length of the paths%
\end{theorem}

\begin{proof}
    By the discussion above we get that  in one of the $|\paths|^{O\left(\frac{a^2 \cdot M^3}{\epsilon^2}\right)}$ loops of the while loop we will find a feasible $F$ satisfying the theorem. As for the running time, we need to compute the running time of checking whether we have an $\epsilon$-Nash flow and whether we have better total cost.

    To check whether we have an $\epsilon$-Nash flow we first consider the network induced by the flow carrying edges, which is a DAG, and check if the longest and the shortest path costs, say $c_{min}$ and $c_{max}$, differ by no more than $\epsilon$, i.e., $c_{max}-c_{min}\leq \epsilon$. Then we consider the full network and check whether the shortest path cost is  $\geq c_{max}-\epsilon$. These can be done in time $O\big(k\cdot poly(n)\big)$ and suffice to verify that we have an  $\epsilon$-Nash flow. Since $k\leq |\paths|$  we get the $|\paths|^{O\left(\frac{a^2 \cdot M^3}{\epsilon^2}\right)}\cdot O\big(poly(n)\big)$ running time. 
\end{proof}



\section{
Instances with Nonincreasing Functions}

In this section  we focus on \flc instances on undirected networks with nonincreasing cost functions where additionally we require  $x \cdot \ell_e(x)$ to be  increasing and concave. For ease of presentation we will call such functions \emph{good}. 
As we show, in the optimal solution demands do not split and thus our approach works also for unsplittable demand. Here, we assume a common opening cost $B_i=B$ for any facility $i$.


We start the section by revealing the structure of the optimal solution (Theorem \ref{theorem:1.3}), which is needed for our algorithm's analysis but will also be used to show that this \flc variant generalizes the Cost-Distance problem, in the sense that solving the Cost-Distance problem reduces to solving \flc for good cost functions (Theorem \ref{theorem:1.4}). Next we present an algorithm for approximately solving \flc for good cost functions and we close the section with the algorithm's analysis, providing approximation guarantees (Theorem \ref{theorem:logS}). 

\subsubsection{The Optimal Solution  is a Forest}
\label{app:isForest}

A key property on which we base our algorithm is that there exists an optimal solution where the flows form a forest on the graph, i.e., there is an optimal solution that opens $k$ facilities and the edges used by the flows do not form cycles and thus  can be seen as $k$ trees rooted in each one of the opened facilities. We prove this in the following lemmas. 

\begin{lemma}
\label{lemma:1.1}
    In any \flc instance with good cost functions, there exists an optimal flow where there are no directed flow carrying cycles, i.e., cycles  in which all demand flows in one direction.
\end{lemma}

\begin{proof}Let $f^*$ be an optimal flow. For any cycle $C$ that  carries positive flow in the same direction, we can decrease $f^*$ on the edges of the cycle by $\min_{e\in C}f^*_e$ getting a feasible solution with cost no more than the initial cost, since  the $x\cdot\ell_e(x)$'s are nondecreasing, and thus with optimal cost. Repeatedly we can eliminate all such cycles and get an optimal acyclic flow. 
\end{proof}


\begin{lemma}
\label{lemma:1.1,5}
    In any \flc instance with good cost functions, there exists an optimal flow where demands do not split their flows once they have met.
\end{lemma}
\begin{proof}

Suppose there is a node in an optimal solution where some demand arrives and then a portion $x_1$ of this demand is routed through path $P_1$ and another portion $x_2$ is routed through path $P_2$ where $P_1 \neq P_2$.  We will show that we can get another optimal flow on the same facilities where $x_1$ and $x_2$ are routed on the same path. Recall that any optimal solution  minimizes the sum $\sum_{e \in E} x_e(F) \cdot \ell_e(x_e(F)) + \sum_{i\in F} B_i$ and thus $\sum_{e \in E} x_e(F) \cdot l_e(x_e(F))$ itself should be minimum among all flows routing to the facilities in $F$. 

If we treat $x_1$ and $x_2$ as variables the total cost that is being paid on the edges of path $P_1$ that are not in $P_2$ and the total cost that is being paid on the edges of path $P_2$ that are not in $P_1$ are respectively: $ \,\,\,C_1(x_1) = \sum_{e \in P_1\setminus P_2} (w'_e+x_1) \cdot l_e(w'_e+x_1)$ and $C_2(x_2) = \sum_{e \in P_2\setminus P_1} (w'_e+x_2) \cdot l_e(w'_e+x_2)\,\,\,$ where $w'_e$ is the rest of the demand that is being routed through $e$ in the optimal solution at hand.

Since all parts of the sum (i.e., the  $(w'_e+x_1) \cdot l_e(w'_e+x_1)$'s) are concave functions (with respect to $x_1$  or $x_2$) then also $C_1(x_1)$ and $C_2(x_2)$ are concave functions.  Since we have assumed that  the flow is optimal, if we take the first derivatives of $C_1(x_1)$ and $C_2(x_2)$ it should hold that $C'_1(x_1)=C'_2(x_2)$. 
 If this was not the case and for example it was $C'_1(x_1)>C'_2(x_2)$ then we could move a small amount of demand from $P_1$ to $P_2$ and the total cost would drop.

 Using $C'_1(x_1)=C'_2(x_2)$ and that $C_1(x_1)$ and $C_2(x_2)$ are concave functions,   we get that for any $d\leq x_1$
$$\frac{C_1(x_1)-C_1(x_1-d)}{d}\geq C'_1(x_1)=C'_2(x_2)\geq \frac{C_2(x_2+d)-C_2(x_2)}{d}$$
\noindent which gives 
$$ C_1(x_1)+C_2(x_2)\geq C_1(x_1-d)+C_2(x_2+d)$$
\noindent implying, by setting $d=x_1$, that we can move all the demand $x_1$  from $P_1$ to $P_2$ without increasing  the total cost, remaining at optimal routing cost.
%
We can repeatedly do this for any split of demand to $P_1$ and $P_2$ ending up with a flow with optimal cost where the demands route their flows on single paths. 
\end{proof}


\begin{lemma}
\label{lemma:1.2}
    In any \flc instance with  good cost functions, there exists an optimal flow for which 
    there are no cycles formed by the edges with positive flow.
\end{lemma}

\begin{proof}
Let $f^*$ be an acyclic optimal flow where the demands do not split, like the one guaranteed by Lemmas \ref{lemma:1.1} and \ref{lemma:1.1,5}. Consider the flow carrying edges and, in order to reach a contradiction, assume that 
there is a cycle formed. Since $f^*$ is acyclic there must be a node $v$ where both of the edges of $v$ send flow away from $v$, contradicting  that in $f^*$ the demands do not split. 
\end{proof}

From Lemmas \ref{lemma:1.1}-\ref{lemma:1.2} it follows that:

\begin{theorem}
\label{theorem:1.3}
     In any \flc instance with  good cost functions, there exists an optimal flow for which 
     there are no cycles formed by the edges with positive flow and the demands use single paths and do not split once they have met.
\end{theorem}

Next we show  that \flc generalizes the Cost-Distance Problem:

\begin{theorem}
\label{theorem:1.4}
    The  Cost-Distance problem polynomially reduces to solving \flc with good cost functions.
\end{theorem}

\begin{proof}
Let $G(V,E)$ be the graph, $S$  be the set of source nodes, $t$ be the target node, $\{w_s\}_{s\in S}$ be the demands of the nodes in $S$ and  $\{c_e\}_{e\in E}$ and $\{l_e\}_{e\in E}$ be the costs and the lengths of the Cost-Distance problem instance, say $\icd$, and recall we are searching for a subgraph $G'$ under which the sum
$\sum_{e\in E'}c_e+\sum_{s_i\in S} w_s\cdot l(s_i,t)$ is minimized, where $l(s_i,t)$ denotes the length of the shortest $s_i$-$t$ path in  $G'$.

To create the \flc instance from the Cost-Distance instance $\icd$ we first note that in a solution  of $\icd$ the total cost incurred by one edge under the corresponding flow $f$  is $c(e) +l(e) \cdot \sum_{s:e \in P_s} w_s$ where $P_s$ is the route chosen for demand $s \in S$ in $f$. We can achieve the same cost by using the following cost function in the \flc instance we create:

\begin{equation*}
    \ell_e(x) = 
     \begin{cases}
       \frac{c(e)}{x}+l(e) &\quad\text{if } x_e  \geq \min_{s \in S}(w_s)\\
       \frac{c(e)}{\min_{s \in S}(w_s)}+l(e) &\quad\text{if } x_e < \min_{s \in S}(w_s)
     \end{cases}
\end{equation*}

\noindent If $x_e=0$ then $x_e\cdot \ell_e(x_e)=0$. If $x_e\geq \min_{s \in S}(w_s)$ then $x_e\cdot \ell_e(x_e)=c(e) +x_e\cdot  l(e) $. Note that for $x \geq \min_{s \in S}(w_s)$ these functions are good. For $x < \min_{s \in S}(w_s)$ we do not really care because no flow less than $\min_{s \in S}(w_s)$ can exist in the optimal solution (recall Theorem \ref{theorem:1.3}) and we just want $\ell_e(0)$ to be bounded so as to pay zero cost when no flow passes through $e$.

The instance of \flc we create has the same graph $G(V,E)
$ with cost functions as defined above, a sufficiently (polynomially) large common facility opening cost $B$, and set of source nodes the set $S$ of $\icd$ augmented by node $t$, i.e., $S\cup\{t\}$. The demands for the nodes in $S$  are as in $\icd$ and for node $t$  is set equal to  the sum of weights of all other demands, i.e., $w_t = \sum_{s \in S} w_s$. $B$ is chosen sufficiently large so that we know that only one facility opens in the optimal solution, since opening a second facility would cost more than the routing costs.

Consider an optimal solution for the \flc instance.  If it opens the facility in a node $v$ different from $t$ then we may get a new optimal solution by moving the facility on $t$, where $w_t$ is already served without any routing cost, and route all the $w_i$'s to $v$ and then to $t$ using the path that $t$ used to send its demand to $v$. This will not increase the cost of the solution since routing a demand of weight  $\sum_{s \in S} w_s$ from $t$ to $v$ (the demand of $t$), is equal to routing a demand of weight $\sum_{s \in S} w_s$ from $v$ to $t$ (the demands of nodes in $S$). Thus we may always change the optimal solution of \flc into one where the facility opens at $t$. 

Given a solution of the \flc instance that opens a facility on $t$, the solution we return for $\icd$ is the subgraph $G(V',E')$ used by the flow of the \flc solution. The optimality of the solution is guaranteed by the fact that minimizing $$\sum_{e \in E} x_e(F) \cdot l_e(x_e(F)) + \sum_{i\in F} B$$ reduces to minimizing $$\sum_{e \in E} x_e(F) \cdot l_e(x_e(F)),$$ since only one facility opens, which, by recalling the definition of the cost functions and the discussion following it, is equal to $$\sum_{e \in E'} [c(e) +x_e(F)\cdot  l(e)],$$ which in turn, by rearranging terms, recalling that demands do not split and recalling that the facility is opened at $t$, is equal to $$\sum_{e\in E'}c_e+\sum_{s_i\in S} w_s\cdot l(s_i,t),$$
\noindent i.e., the objective of the Cost-Distance problem. 
\end{proof}

\subsubsection{The Algorithm}
\label{sec:algoDec}

An observation that simplifies our approach  is that we may focus on the $k$-median variant of \flc. In this variant there are no opening costs for the facilities, only routing costs, and we may open at most $k$ facilities. 
 Solving the $k$-median variant for all possible $k$ values,  solves the original problem since for any $k$ we can compute the sum of the routing cost of the  optimal solution  with the original problem's opening cost, which for any $k$  will be $k\cdot B$ (recall, we have a common cost $B_i=B$) and then compare these sums for the $n$ different  values of $k$ in order to keep the optimal one. 
 

Below we give an algorithm that takes as input an instance of \flc and returns $k$ facility points for the k-median version of \flc. The algorithm is influenced by the algorithm proposed by Meyerson et al. \cite{Bicriteria} for the Cost-Distance problem which returns one target point (i.e., one facility, in our setting). However, their matching mechanism and analysis are tailored made to the fact that the total cost incurred by each edge is linear. We introduce a novel matching mechanism which in turn complicates the resulting analysis. For the algorithm we will need the following distance metric: $g(u,v,w) = \sum_{e \in P_{u,v}^w} w \cdot \ell_e(w)$ where $P_{u,v}^w$ is the closest path from $u$ to $v$ with respect to the distance metric $\ell_e(w)$. 

To give a high level view of the algorithm, the algorithm works in phases. In every phase it pairs up the demands. For every such pair it routes both demands  from their sources to a chosen meeting point, and then it randomly (in a biased way) routes them back to one of the two demand sources. For the next phase, every pair of paired demands is handled as one demand of volume equal to the sum of the two demands and source the randomly chosen source in the previous phase. The algorithm will terminate when some phase ends with $k$ demands, and it will open facilities right on the sources of these demands. 

Two important ingredients of the algorithm is how the pairing is done (steps 2(a) and 2(b) below) and how the common source for the merging demands is chosen (step 2(d)ii).  The reasons for doing the pairing and the routing this way will become apparent in the next section that analyzes the algorithm.  For step 2(b) the requirement for having nodes unmatched is there to avoid forcing to match nodes that are in different rooted trees of the optimal solution (in case these are of odd cardinality). Recall that the optimal solution can be seen as $k$ trees rooted in each one of the opened facilities.

\bigskip
\noindent\textbf{Approximation scheme for nonincreasing cost functions}
\begin{enumerate}
    \item Initialize $S_0 = S$, $w_{0,s} = w_s$ and $i = 0$.
    \item While $|S_i|>k$:
    \begin{enumerate}
        \item For every pair $u,v \in S_i$ find $K_i(u,v) = \min_{z \in V}\{g(u,z,w_{i,u}) + g(v,z,w_{i,v}) \\ + \frac{w_{i,u}}{w_{i,u}+w_{i,v}} \cdot g(z,u,w_{i,u}+w_{i,v}) + \frac{w_{i,v}}{w_{i,u}+w_{i,v}} \cdot g(z,v,w_{i,u}+w_{i,v})\}$.
        \item Perform a maximum cardinality minimum cost matching  on $S_i$ with respect to the costs $K_i$
        under the constraint that, unless you get an empty matching, 
        $k$ nodes must be unmatched. 
        \item Set $S_{i+1} = \{\}$.
        \item For each matched $u,v$ pair:
        \begin{enumerate}
            \item Send both demands to the node $z$ that minimized the expression of step 2(a).
            \item Choose $u$ with probability $\frac{w_{i,u}}{w_{i,u}+w_{i,v}}$, otherwise chose $v$. Without loss of generality we will assume that we chose $u$.
            \item Send the combined $w_{i,u}+w_{i,v}$ demand back to $u$, add $u$ to $S_{i+1}$ and set $w_{i+1,u} = w_{i,u}+w_{i,v}$.
        \end{enumerate}
        \item Add unmatched nodes to $S_{i+1}$
        \item Set $i \leftarrow i+1$
    \end{enumerate}
    \item Return as facilities the $k$ nodes that are in $S_i$ and as flows the ones dictated by the above procedure.
\end{enumerate}

\subsubsection{The Analysis}

For our analysis we introduce the metric $C_u^i$ to be the total cost payed due to the movement of demand from source $u$ in phase $i$. The total cost payed in phase $i$ is $C^i = \sum_{s \in S_i} C_s^i$. Since at every step the number of sources minus $k$ is roughly divided by two and we end up with $k$ (final) sources, it is not difficult to show the following lemma.

\begin{lemma}
\label{lemma:1.5}
    The algorithm presented terminates after $O(\log|S|)$ phases. 
\end{lemma}

If we prove that in each phase of the algorithm the expected cost payed (i.e. $\mathbf{E}[C^i]$ is at most the cost of the optimal solution, then combining this fact with Lemma \ref{lemma:1.5} shows that our algorithm is an $O(\log|S|)$-approximate algorithm for the \flc with good functions. We will 
show that this is true for the first phase and 
that the expected cost of each phase is less than or equal to the cost of the first phase.

\subsubsection{Phase 0}

Phase 0 is the first phase of our algorithm where no demands have been aggregated yet. We begin with the following lemma which is a generalization of \cite[Lemma 4.2]{Bicriteria} and reveals some structural property of the optimal solution. Recall, again, that the optimal solution can be seen as $k$ trees rooted in each one of the opened facilities.

\begin{lemma}
\label{lemma:1.6}
    Let $S' \subseteq S$ be a set of sources that are routed at a facility in node $r$ in the optimal solution. Lets call $T$ 
    the tree that corresponds to the flows from nodes in $S'$ to $r$. Then there exists a matching like the one done in steps 2(a),(b) of our algorithm where 
    demands     only use edges that they use in the optimal flow and at most one source is left unmatched.
\end{lemma}

\begin{proof}
Intuitively, we perform steps 2(a),(b) conditioned that the paths for computing the  $g(\cdot,\cdot,\cdot)$'s are restricted to belong in $T$. For every $s \in S'$ we take its path towards $r$ at $T$ until it meets with the path of another source. We will call the vertex in which the two paths merge a level one meeting point. We move all sources at their corresponding level one meeting points. In each meeting point with an even number of demands we match them arbitrarily until no demand is left unmatched. In each meeting point with an odd number of demands we do the same thing however now there will be one demand left out. We continue along the path of those level one meeting points with an unmatched demand until their path merges with the path of another level one meeting point. The vertices in which those level one paths meet are called level two meeting points. We continue along the same line of thought until we reach our final meeting point $r$ where at most one demand will be left unmatched. 
\end{proof}

In the algorithm in steps 2(a)-(b)  we search for the best (w.r.t. cost) such matching and we choose as $z$ (in step 2(d)i) the respective meeting point described by the proof of Lemma \ref{lemma:1.6}. Thus, the returned matching cannot be worse than the one dictated by Lemma \ref{lemma:1.6} and it will be without loss of generality that in this phase but also in every other phase we can assume that the edges used are exactly those found by the matching (we will route the same demand on paths at most as expensive, in total, as those of the optimal solution). To bound the cost of our matching we just need to bound the cost of sending demands to those meeting points and the expected demand of sending them back again. The first half is accomplished with the following lemma.

\begin{lemma}
\label{lemma:1.7}
    The cost of sending all demands to their respective meeting points is less than that of the optimal solution.
\end{lemma}

\begin{proof}
From Lemma \ref{lemma:1.6} there is a subtle implication that becomes very useful right now. Each edge is traversed by demands that traverse that edge in the optimal solution. More specifically suppose that an edge $e \in E$ is traversed by a set $S''$ of demands. In our matching only one of this demands traverses that edge. Thus the amount of flow $f'_e$ traversing edge $e$ in our matching is less or equal than the amount of flow $f_e$ that traverses $e$ in the optimal solution. And since $x \cdot \ell_e(x)$ is nondecreasing we have that $f'_e \cdot \ell_e(f'_e) \leq f_e \cdot \ell_e(f_e)$. Summing over all edges we get that the total cost is less than the optimal cost. 
\end{proof}

The next Lemma holds for any phase $i$ of the algorithm.

\begin{lemma}
\label{lemma:1.8}
    The expected cost of routing demands back from their meeting point to the selected source is less than or equal to the cost paid for sending them to the meeting point.
\end{lemma}

\begin{proof}
Suppose we are at phase $i$ and let  $a_1,a_2,\ldots,a_m$ be the edges along the path of $u$ to the meeting point and $b_1,b_2,\ldots,b_n$ the ones of $v$. Then the expected cost $\mathbf{E}[C_{\text{back}}]$ of sending them back to a source is the cost of sending them to $u$ times the probability of choosing $u$ plus the probability of sending them to $v$ times the cost of sending them there. In other words we have that $$\mathbf{E}[C_{\text{back}}] = (w_{i,u}+w_{i,v}) \cdot \frac{w_{i,u}}{w_{i,u}+w_{i,v}} \cdot \sum_{j=1}^m\ell_{a_j}(w_{i,u}+w_{i,v})
+(w_{i,u}+w_{i,v}) \cdot \frac{w_{i,v}}{w_{i,u}+w_{i,v}} \cdot \sum_{j=1}^n\ell_{b_j}(w_{i,u}+w_{i,v})
$$$$\leq w_{i,u} \cdot \sum_{j=1}^m\ell_{a_j}(w_{i,u})
+w_{i,v} \cdot \sum_{j=1}^m\ell_{b_j}(w_{i,v})$$ 
which is exactly the cost of sending those demands to the meeting point. The inequality holds since the demands are positive and the  $\ell_e$'s are nonincreasing.
\end{proof}

From Lemmas \ref{lemma:1.7} and \ref{lemma:1.8} we can see that $C^0 \leq 2 \cdot C^*$ where $C^*$ is the total cost of the optimal solution.  

\subsubsection{Phase $i$}

The only thing left to do is to bound the cost of sending demands to their meeting points in phase $i$ of the algorithm. This can be done using the following lemma.

\begin{lemma}
\label{lemma:1.9}
    The expected cost of routing demands to their meeting points in phase $i$ is less than the cost of the optimal solution.
\end{lemma}

\begin{proof}
Suppose at phase $i$ that in a node $u$ a set $S_u''$ of demands has been gathered with a total demand $W_u = \sum_{s \in S_u''} w_s$. The probability of $u$ having this demand is $\frac{w_u}{W_u}$. This claim is easy to see because for $u$ to have that demand it must have been selected in all phases with a total probability of $$\frac{w_u}{w_u+w_{v_1}} \cdot \frac{w_u+w_{v_1}}{w_u+w_{v_1}+w_{v_2}} \cdot 
\ldots\cdot\frac{\sum_{s \in S_u''} w_s-w_{u_{last}}}{\sum_{s \in S_u''} w_s}$$ and in this product only the first numerator and the last denominator do not cancel out. We once again perform the matching described by Lemma \ref{lemma:1.6} with respect to weights $w_{i,u}$. 

We are now going to show that the expected cost of the matching described by Lemma \ref{lemma:1.6} and searched for in steps 2(a)-(b) is bounded by the cost of the optimal solution. For the sake of the argument lets call $g_e(x) = x \cdot \ell_e(x)$ the cost payed for the usage of edge $e$ (recall $g_e(x)$ is a nondecreasing concave function). Suppose that a specific edge $e$ in the optimal solution is used by demands $w_1, w_2, \ldots,  w_k$. Thus the total cost payed for edge $e$ in the optimal solution is $g_e\left( \sum_{j=1}^k w_j \right)$. We will show that for this arbitrary $e$ it is $\mathbf{E}[C_{i,e}] \leq g_e\left( \sum_{j=1}^k w_j \right)$, where $C_{i,e}$ is the cost payed in the matching of phase $i$ by $e$. Summing over all $e$ will conclude the proof.

In phase $i$ the expected cost payed by edge $e$ because of demand $w_j$ is the cost payed because a total demand of weight $W_{w_j}$ passes through $e$ times the probability that all of the demand with which $w_j$ has been matched, actually passes through $e$. That probability can be expressed as the probability of $W_{w_j}$ ending up in $w_j$ (which we have shown earlier that is equal to $\frac{w_j}{W_{w_j}}$), times the probability that demand is not matched earlier in the Tree. We will call the later probability $p_{e,w_j}$. Thus the expected cost payed by edge $e$ due to demand $w_j$ is $\frac{w_j}{W_{w_j}} \cdot p_{e,w_j} \cdot g_e(W_{w_j})$. Also we will call $p_{e,0}$ the probability that no demand passes through $e$ in our matching either because with probability $\prod_{j=1}^k \left( 1 - \frac{w_j}{W_{w_j}} \right)$ there are no demands in the subtree underneath $e$ or there was an even number of demands in the subtree underneath $e$ and thus all demands have been matched lower in the Tree. So the total expected cost of edge $e$ is $p_{e,0} \cdot g_e(0) + \sum_{j=1}^k \frac{w_j}{W_{w_j}} \cdot p_{e,w_j} \cdot g_e(W_{w_j})$. However, one can see that  $p_{e,0} + \sum_{j=1}^k \frac{w_j}{W_{w_j}} \cdot p_{e,w_j} = 1$ and $g_e(x)$ is concave, so we can use Jensen's inequality \cite{jensen_inequality} for concave functions which leads to the following expression:

\begin{equation*}
    \mathbf{E}[C_{i,e}] \leq p_{e,0} \cdot g_e(0) + \sum_{j=1}^k \frac{w_j}{W_{w_j}} \cdot p_{e,w_j} \cdot g_e(W_{w_j}) 
\end{equation*}
\begin{equation*}
   \leq g_e\left( p_{e,0} \cdot 0 + \sum_{j=1}^k \frac{w_j}{W_{w_j}} \cdot p_{e,w_j} \cdot W_{w_j}  \right)
     = g_e\left( \sum_{j=1}^k w_j \cdot p_{e,w_j}\right)\leq  g_e\left( \sum_{j=1}^k w_j \right)
\end{equation*}

\noindent where the last inequality holds because $p_{e,w_j} \leq 1$ and $g_e(x)$ is nondecreasing. This argument concludes our proof. \end{proof}

Combining the aforementioned lemmas we end up with the following lemma.

\begin{lemma}
\label{lemma:1.10}
    The routing cost of each phase is on expectation at most the cost of the optimal solution.
\end{lemma}

Combining this lemma with the fact that our algorithm has $O(\log|S|)$ phases we get the following theorem.

\begin{theorem}
\label{theorem:logS}
    The analyzed algorithm  produces an $O(\log|S|)$-approximate solution to \flc with good cost functions.
\end{theorem}

The  solution returned by the approximation scheme for nonincreasing cost functions can be used for approximately solving \fls. Under the opened set of facilities, say $\hat{F}$, the routing cost when considering \fls, i.e., when requiring agents to be at equilibrium, is at most $PoA(D)$ times the optimal routing cost. Here, $PoA(D)$ is the Price of Anarchy \cite{PoA}  for cost functions in class $D$ \cite{DBLP:journals/mor/CorreaSM04}, which captures how much worse can an equilibrium flow be when compared to the optimal flow for the cost functions used.

Thus, the total cost of the solution that opens facility set $\hat{F}$ when considering \fls is at most $PoA(D)$ times the total cost of the solution when considering \flc, which in turn is at most $O(\log|S|)$ times the cost of the optimal solution for \flc. On the other, hand the optimal solution for \flc has total cost at most as the total cost of the optimal solution for \fls. Putting it all together we get the following corollary. We note that our scope is not to provide a PoA(D) bound, rather, it is to provide an approximation scheme for when such a bound is known.

\begin{corollary}\label{cor:poa}
    The approximation scheme presented  outputs an $[O(\log|S|)\cdot PoA(D)]$-approximate solution to \fls with good cost functions.
\end{corollary}

\section{Conclusion and the Case of Discrete Agents}

We introduced Facility Location problems that account for congestion dependent connection costs. Among the many possible variants, we presented algorithms for two of them. Providing results for any of these variants is an interesting direction.

Our results can be of use  if one considers discrete agents. One can use the algorithm in Section 3, that cuts the flow in small discrete particles, to solve the corresponding continuous case and then apply some type of rounding to get  approximation guarantees.  These guarantees will be  
close to the ones of the continuous case when the number of agents is large enough to allow for small enough cuts in the flow. In Section 4 we showed that in the optimal solution the demands do
not split. Thus, the algorithm presented here works also for the discrete agents' case. Last, derandomizing this algorithm  could be an interesting extension.

\bibliographystyle{plain}
\bibliography{refs}

@inproceedings{Simple;And;Better,
author = {Gupta, Anupam and Kumar, Amit and Roughgarden, Tim},
title = {Simpler and Better Approximation Algorithms for Network Design},
year = {2003},
isbn = {1581136749},
publisher = {Association for Computing Machinery},
address = {New York, NY, USA},
url = {https://doi.org/10.1145/780542.780597},
doi = {10.1145/780542.780597},
booktitle = {Proceedings of the Thirty-Fifth Annual ACM Symposium on Theory of Computing},
pages = {365–372},
numpages = {8},
location = {San Diego, CA, USA},
series = {STOC '03}
}

@inproceedings{BicriteriaOLD,
  title={Cost-distance: two metric network design},
  author={Meyerson, A and Munagala, K and Plotkin, S},
  booktitle={Proceedings 41st Annual Symposium on Foundations of Computer Science},
  pages={624--624},
  year={2000},
  organization={IEEE Computer Society}
}

@article{Bicriteria,
  author       = {Adam Meyerson and
                  Kamesh Munagala and
                  Serge A. Plotkin},
  title        = {Cost-Distance: Two Metric Network Design},
  journal      = {{SIAM} J. Comput.},
  volume       = {38},
  number       = {4},
  pages        = {1648--1659},
  year         = {2008},
  url          = {https://doi.org/10.1137/050629665},
  doi          = {10.1137/050629665},
  bibsource    = {dblp computer science bibliography, https://dblp.org}
}

@article{chekuri2001deterministic,
  title={A deterministic algorithm for the cost-distance problem},
  author={Chekuri, Chandra and Khanna, Sanjeev and Naor, Joseph},
  journal={Symposium on Discrete Algorithms: Proceedings of the twelfth annual ACM-SIAM symposium on Discrete algorithms},
  volume={7},
  number={09},
  pages={232--233},
  year={2001}
}

@article{UFL:7.88,
  title={Universal Facility Location},
  author={Mohammad Mahdian and Martin Pal},
  journal={Springer, Berlin, Heidelberg},
  year={2003},
  isbn = {978-3-540-39658-1},
  doi = {0.1007/978-3-540-39658-1_38}
}

@phdthesis{Pandit:Disertation,
  title={Local Search Heuristics For Facility Location Problems},
  author={Vinayaka Pandit},
  school={Department of Computer Science and Engineering Indian Institute of Technology Delhi},
  year={2004}
}

@article{UFL:5,
author = {Bansal, Manisha and Garg, Naveen and Gupta, Neelima},
year = {2018},
month = {12},
pages = {},
title = {A 5-Approximation for Universal Facility Location},
journal={ IARCS Annual Conference on Foundations of Software Technology and Theoretical Computer Science (FSTTCS)},
doi = {10.4230/LIPIcs.FSTTCS.2018.24}
}

@article{PoA,
author = {Koutsoupias, Elias and Papadimitriou, Christos},
year = {1999},
pages = {404- 413},
title = {Worst-case equilibria},
journal = {16th Annual Symposium on Theoretical Aspects of Computer Science (STACS)},
doi = {10.1016/j.cosrev.2009.04.003}
}

@article{AlthoferFotakis,
title = {Efficient methods for selfish network design},
journal = {Theoretical Computer Science},
volume = {448},
pages = {9-20},
year = {2012},
issn = {0304-3975},
doi = {https://doi.org/10.1016/j.tcs.2012.04.033},
url = {https://www.sciencedirect.com/science/article/pii/S0304397512003982},
author = {Dimitris Fotakis and Alexis C. Kaporis and Paul G. Spirakis}
}

@article{ApproxCaratheodoryBarman,
  author       = {Siddharth Barman},
  title        = {Approximating Nash Equilibria and Dense Subgraphs via an Approximate
                  Version of Carath{\'{e}}odory's Theorem},
  journal      = {{SIAM} J. Comput.},
  volume       = {47},
  number       = {3},
  pages        = {960--981},
  year         = {2018},
  url          = {https://doi.org/10.1137/15M1050574},
  doi          = {10.1137/15M1050574},
  bibsource    = {dblp computer science bibliography, https://dblp.org}
}

@inproceedings{Hierarchical,
author = {Guha, S. and Meyerson, A. and Munagala, K.},
title = {Hierarchical Placement and Network Design Problems},
year = {2000},
isbn = {0769508502},
publisher = {IEEE Computer Society},
address = {USA},
booktitle = {Proceedings of the 41st Annual Symposium on Foundations of Computer Science},
pages = {603},
series = {FOCS '00}
}

@article{JainVazirani,
author = {Jain, Kamal and Vazirani, Vijay},
year = {2001},
month = {03},
title = {Approximation algorithms for metric facility location and k-Median problems using the primal-dual schema and Lagrangian relaxation},
volume = {48},
journal = {Journal of The ACM - JACM},
doi = {10.1145/375827.375845}
}

@article{jensen_inequality, 
title={Sur les fonctions convexes et les inégalités entre les valeurs moyennes}, 
volume={30}, 
DOI={10.1007/bf02418571}, 
journal={Acta Mathematica}, 
author={Jensen, J. L. W. V.}, 
year={1906}, 
pages={175–193}
}

@book{FLbook1_Man_Science, 
title={ Facility Location.
Concepts, Models, Algorithms and Case Studies}, 
DOI={10.1007/978-3-7908-2151-2}, 
publisher={Springer-Verlag Berlin}, 
author={Reza Zanjirani Farahani, Masoud Hekmatfar}, 
year={2009}
}

@book{FLbook2_ThAndApp, 
title={Facility Location. Applications and Theory}, 
publisher={Springer-Verlag Berlin}, 
author={Zvi Drezner, Horst W. Hamacher}, 
year={2001}
}

@article{li_2013,
title = {A 1.488 approximation algorithm for the uncapacitated facility location problem},
journal = {Information and Computation},
volume = {222},
pages = {45-58},
year = {2013},
note = {38th International Colloquium on Automata, Languages and Programming (ICALP 2011)},
issn = {0890-5401},
doi = {https://doi.org/10.1016/j.ic.2012.01.007},
url = {https://www.sciencedirect.com/science/article/pii/S0890540112001459},
author = {Shi Li}
}

@inproceedings{shmoys_tardos_aardal_1997,
author = {Shmoys, D. B. and Tardos, E. and Aardal, K.I.},
title = {Approximation algorithms for facility location problems},
year = {1997},
booktitle = {Proceedings of 29th Annual ACM Symposium on Theory of Computing},
pages = {265-274}
}

@article{chudak_shmoys_2003, title={Improved Approximation Algorithms for the Uncapacitated Facility Location Problem}, volume={33}, DOI={10.1137/s0097539703405754}, number={1}, journal={SIAM Journal on Computing}, author={Chudak, Fabián A. and Shmoys, David B.}, year={2003}, pages={1–25}}

@article{guha_khuller_1999, title={Greedy Strikes Back: Improved Facility Location Algorithms}, volume={31}, DOI={10.1006/jagm.1998.0993}, number={1}, journal={Journal of Algorithms}, author={Guha, Sudipto and Khuller, Samir}, year={1999}, pages={228–248}}

@inproceedings{charikar_chekuri_goel_guha_1998,
author = {Charikar, Moses and Chekuri, Chandra and Goel, Ashish and Guha, Sudipto},
title = {Rounding via Trees: Deterministic Approximation Algorithms for Group Steiner Trees and k-Median},
year = {1998},
isbn = {0897919629},
publisher = {Association for Computing Machinery},
address = {New York, NY, USA},
url = {https://doi.org/10.1145/276698.276719},
doi = {10.1145/276698.276719},
booktitle = {Proceedings of the Thirtieth Annual ACM Symposium on Theory of Computing},
pages = {114–123},
numpages = {10},
location = {Dallas, Texas, USA},
series = {STOC '98}
}

@article{charikar_guha_tardos_shmoys_2002,
title = {A Constant-Factor Approximation Algorithm for the k-Median Problem},
journal = {Journal of Computer and System Sciences},
volume = {65},
number = {1},
pages = {129-149},
year = {2002},
issn = {0022-0000},
doi = {https://doi.org/10.1006/jcss.2002.1882},
url = {https://www.sciencedirect.com/science/article/pii/S0022000002918829},
author = {Moses Charikar and Sudipto Guha and Éva Tardos and David B. Shmoys},
}

@article{bartal_1996, title={Probabilistic approximation of metric spaces and its algorithmic applications}, DOI={10.1109/sfcs.1996.548477}, journal={Proceedings of 37th Conference on Foundations of Computer Science}, author={Bartal, Y.}, year={1996}}

@article{bartal_1998, title={On approximating arbitrary metrices by tree metrics}, DOI={10.1145/276698.276725}, journal={Proceedings of the thirtieth annual ACM symposium on Theory of computing - STOC 98}, author={Bartal, Yair}, year={1998}}

@article{arya_garg_khandekar_meyerson_munagala_pandit_2004, title={Local Search Heuristics for k-Median and Facility Location Problems}, volume={33}, DOI={10.1137/s0097539702416402}, number={3}, journal={SIAM Journal on Computing}, author={Arya, Vijay and Garg, Naveen and Khandekar, Rohit and Meyerson, Adam and Munagala, Kamesh and Pandit, Vinayaka}, year={2004}, pages={544–562}}

@article{jain_mahdian_markakis_saberi_vazirani_2003, title={Greedy facility location algorithms analyzed using dual fitting with factor-revealing LP}, volume={50}, DOI={10.1145/950620.950621}, number={6}, journal={Journal of the ACM}, author={Jain, Kamal and Mahdian, Mohammad and Markakis, Evangelos and Saberi, Amin and Vazirani, Vijay V.}, year={2003}, pages={795–824}}

@article{aardal_chudak_shmoys_1999, title={A 3-approximation algorithm for the k-level uncapacitated facility location problem}, volume={72}, DOI={10.1016/s0020-0190(99)00144-1}, number={5-6}, journal={Information Processing Letters}, author={Aardal, Karen and Chudak, Fabián A. and Shmoys, David B.}, year={1999}, pages={161–167}}

@article{karger_minkoff_2000,
author = {Karger, David and Minkoff, Michael},
year = {2000},
month = {02},
pages = {613-623},
title = {Building Steiner Trees with Incomplete Global Knowledge.},
journal = {Annual Symposium on Foundations of Computer Science - Proceedings},
doi = {10.1109/SFCS.2000.892329}
}

@article{gupta_kleinberg_kumar_rastogi_yener_2001,
author = {Gupta, Anupam and Kleinberg, Jon and Kumar, Amit and Rastogi, Rajeev and Yener, Bulent},
year = {2001},
month = {01},
pages = {389-398},
title = {Provisioning a virtual private network: A network design problem for multicommodity flow},
journal = {33rd Proc. STOC},
doi = {10.1145/380752.380830}
}

@article{swamy_kumar_2004,
author = {Swamy, Chaitanya and Kumar, Amit},
year = {2004},
month = {01},
pages = {245-269},
title = {Primal–Dual Algorithms for Connected Facility Location Problems},
volume = {40},
journal = {Algorithmica},
doi = {10.1007/s00453-004-1112-3}
}

@article{vygen_2007,
author = {Vygen, Jens},
year = {2007},
month = {07},
pages = {427-433},
title = {From stars to comets: Improved local search for universal facility location},
volume = {35},
journal = {Oper. Res. Lett.},
doi = {10.1016/j.orl.2006.08.004}
}

@article{angel_thang_regnault_2014,
author = {Angel, Eric and Thang, Nguyen and Regnault, Damien},
year = {2014},
month = {02},
pages = {},
title = {Improved Local Search for Universal Facility Location},
volume = {29},
isbn = {978-3-642-38767-8},
journal = {Journal of Combinatorial Optimization},
doi = {10.1007/978-3-642-38768-5_29}
}

@article{chuzhoy_gupta_noar_sinha_2008,
author = {Julia Chuzhoy  and Anupam  Gupta and  Joseph (Seffi) Naor  and mitabh Sinha},
title = {On the Approximability of Some Network Design Problems},
year = {2008},
issue_date = {May 2008},
publisher = {Association for Computing Machinery},
address = {New York, NY, USA},
volume = {4},
number = {2},
issn = {1549-6325},
url = {https://doi.org/10.1145/1361192.1361200},
doi = {10.1145/1361192.1361200},
journal = {ACM Trans. Algorithms},
month = may,
articleno = {23},
numpages = {17}
}

@article{Rosenthal1973,
author = {Rosenthal, Robert W.},
year = {1973},
month = {12},
pages = {},
title = {A class of games possessing pure-strategy Nash equilibria},
volume = {7},
isbn = {978-3-540-22339-9},
journal = {International Journal of Game Theory},
doi = {10.1007/BF01737559}
}

@article{FotLianDimSer_caraththoedo,
  author       = {Sotirios Dimos and
                  Dimitris Fotakis and
                  Thanasis Lianeas and
                  Kyriakos Sergis},
  title        = {Escaping Braess's paradox through approximate Caratheodory's theorem},
  journal      = {Inf. Process. Lett.},
  volume       = {179},
  pages        = {106289},
  year         = {2023},
  url          = {https://doi.org/10.1016/j.ipl.2022.106289},
  doi          = {10.1016/J.IPL.2022.106289},
  bibsource    = {dblp computer science bibliography, https://dblp.org}
}

@article{DBLP:journals/mor/CorreaSM04,
  author       = {Jos{\'{e}} R. Correa and
                  Andreas S. Schulz and
                  Nicol{\'{a}}s E. Stier Moses},
  title        = {Selfish Routing in Capacitated Networks},
  journal      = {Math. Oper. Res.},
  volume       = {29},
  number       = {4},
  pages        = {961--976},
  year         = {2004},
  url          = {https://doi.org/10.1287/moor.1040.0098},
  doi          = {10.1287/MOOR.1040.0098},
  bibsource    = {dblp computer science bibliography, https://dblp.org}
}

@article{Bra68,
    title   = "{\"{U}ber ein paradox aus der Verkehrsplanung}",
    author  = {D. Braess},
    journal = {Unternehmensforschung},
    volume  = {12},
    year    = {1968},
    pages   = {258--268},
}

@article{Mur70,
 author = {J. D. Murchland},
 title = {Braess's paradox of traffic flow},
 journal = {Transportation Res.},
 volume = {4},
 year = {1970},
 pages = {391--394},
 }

@article{BraessUnsolvable,
    title = "On the severity of Braess's Paradox: Designing networks for selfish users is hard",
    journal = "Journal of Computer and System Sciences",
    volume = "72",
    number = "5",
    pages = "922 - 953",
    year = "2006",
    note = "Special Issue on FOCS 2001",
    issn = "0022-0000",
    doi = "https://doi.org/10.1016/j.jcss.2005.05.009",
    url = "http://www.sciencedirect.com/science/article/pii/S0022000006000109",
    author = "Tim Roughgarden",
}

@Book{Kel08,
    title         = "{The mathematics of traffic in networks}",
    author        = {F. Kelly},
    publisher     = {In {\em The Princeton Companion to Mathematics} (Editors: T. Gowers, J. Green and I. Leader).  Princeton University Press},
    year          = {2008},
}

@Book{RouB05,
    title         = "{Selfish Routing and the Price of Anarchy}",
    author        = {T. Roughgarden},
    publisher     = {MIT press},
    year          = {2005},
}

@article{ValiantR10,
  author    = {Gregory Valiant and
               Tim Roughgarden},
  title     = {Braess's Paradox in large random graphs},
  journal   = {Random Struct. Algorithms},
  volume    = {37},
  number    = {4},
  pages     = {495--515},
  year      = {2010}
}

@InProceedings{CY10,
author 	    = {Chung, Fan and Young, Stephen J.},
title       = {Braess's Paradox in Large Sparse Graphs},
booktitle   = {Internet and Network Economics},
year        = {2010},
publisher   = {Springer Berlin Heidelberg},
address     = {Berlin, Heidelberg},
pages       = {194--208}
}

@article{SeifFacLocCong,
author = {Seifbarghy, Mehdi and Mansouri, Aida},
year = {2016},
month = {01},
pages = {281},
title = {Modelling and solving a congested facility location problem considering systems' and customers' objectives},
volume = {22},
journal = {International Journal of Industrial and Systems Engineering},
doi = {10.1504/IJISE.2016.074707}
}

@article{EshaghiFacLocCong,
author = {Eshaghi, Amir and Jahani, Hamed and Aghaie, Abdollah and Ivanov, Dmitry},
year = {2019},
month = {01},
pages = {2279-2284},
title = {A multi-layer congested facility location problem with consideration of impatient customers in a queuing system},
volume = {52},
journal = {IFAC-PapersOnLine},
doi = {10.1016/j.ifacol.2019.11.545}
}

@unknown{ChakrabortFacLocCong,
author = {Chakraborty, Arghya and Vaze, Rahul},
year = {2022},
month = {11},
pages = {},
title = {Online facility location with timed-requests and congestion},
doi = {10.48550/arXiv.2211.11961}
}

@article{WANG2023109167,
title = {A multi-objective fuzzy facility location problem with congestion and priority for drone-based emergency deliveries},
journal = {Computers \& Industrial Engineering},
volume = {179},
pages = {109167},
year = {2023},
issn = {0360-8352},
doi = {https://doi.org/10.1016/j.cie.2023.109167},
url = {https://www.sciencedirect.com/science/article/pii/S0360835223001912},
author = {Xin Wang and Jiemin Zhao and Chun Cheng and Mingyao Qi},

}

@article{JALILIMARAND2024442,
title = {A congested facility location problem with strategic customers},
journal = {European Journal of Operational Research},
volume = {318},
number = {2},
pages = {442-456},
year = {2024},
issn = {0377-2217},
doi = {https://doi.org/10.1016/j.ejor.2024.05.026},
url = {https://www.sciencedirect.com/science/article/pii/S0377221724003825},
author = {Ata {Jalili Marand} and Pooya Hoseinpour}
}

@article{DBLP:journals/msom/BaronBK08,
  author       = {Opher Baron and
                  Oded Berman and
                  Dmitry Krass},
  title        = {Facility Location with Stochastic Demand and Constraints on Waiting
                  Time},
  journal      = {Manuf. Serv. Oper. Manag.},
  volume       = {10},
  number       = {3},
  pages        = {484--505},
  year         = {2008},
  url          = {https://doi.org/10.1287/msom.1070.0182},
  doi          = {10.1287/MSOM.1070.0182},
  bibsource    = {dblp computer science bibliography, https://dblp.org}
}

@article{DBLP:journals/ior/DanM19,
  author       = {Teodora Dan and
                  Patrice Marcotte},
  title        = {Competitive Facility Location with Selfish Users and Queues},
  journal      = {Oper. Res.},
  volume       = {67},
  number       = {2},
  pages        = {479--497},
  year         = {2019},
  url          = {https://doi.org/10.1287/opre.2018.1781},
  doi          = {10.1287/OPRE.2018.1781},
  bibsource    = {dblp computer science bibliography, https://dblp.org}
}

@article{MARIANOV200832,
title = {Facility location for market capture when users rank facilities by shorter travel and waiting times},
journal = {European Journal of Operational Research},
volume = {191},
number = {1},
pages = {32-44},
year = {2008},
issn = {0377-2217},
doi = {https://doi.org/10.1016/j.ejor.2007.07.025},
url = {https://www.sciencedirect.com/science/article/pii/S0377221707008685},
author = {Vladimir Marianov and Miguel Ríos and Manuel José Icaza}
}

@article{HajiaghayiMM03,
  author       = {Mohammad Taghi Hajiaghayi and
                  Mohammad Mahdian and
                  Vahab S. Mirrokni},
  title        = {The facility location problem with general cost functions},
  journal      = {Networks},
  volume       = {42},
  number       = {1},
  pages        = {42--47},
  year         = {2003},
  url          = {https://doi.org/10.1002/net.10080},
  doi          = {10.1002/NET.10080}
}



\appendix


\section{On the Applicability of Facility Location Techniques}
\label{app:FLtechniquesFail}

\subsection{Why Local Search does not Work in our Setting}

One natural idea would be to try and solve this problem using local search algorithms. 
Consider the following example for instances where the cost function  of an edge is a polynomial of bounded degree $d$. Since in our problem we have constant facility costs we are going to examine local steps that do not take into consideration the amount of demand each facility accommodates. We are going to use the $\textbf{open}$ local step where one facility is opened and the demand is rerouted optimally, the $\textbf{close}$ local step where an open facility is closed and the demand is rerouted optimally and the $\textbf{swap}$ where an opened facility is closed while a closed facility is opened simultaneously and the demand is the rerouted optimally.

\begin{figure}[ht]
\centerline{\includegraphics[scale=.34]{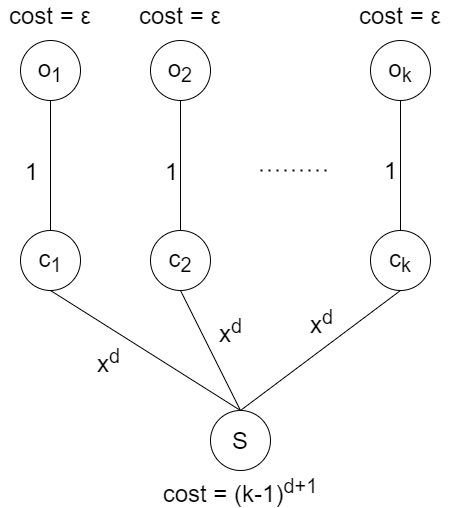}}
\caption{A counter example for the use of Local Search}
\label{fig:localSearch}
\end{figure}

In the example illustrated in figure \ref{fig:localSearch} we have nodes $c_1, c_2, ..., c_k$ all having unit demand. The cost of opening a facility on $o_i$ is $\epsilon$ and the cost functions of edges $(c_i, o_i)$ are constant and equal to $1$. We also have another possible facility node $S$ with cost of opening a facility $(k-1)^{d+1}$. The cost functions of edges $e = (c_i, S)$ is $l_e(x) = x^d$. The optimal solution is opening all $o_i$ facilities with cost $k \cdot \epsilon$ and route each demand $c_j$ to its corresponding facility $o_j$. Now consider the following locally optimal solution where only $S$ is open. Obviously we cannot close $S$. We cannot open $o_i$ because it would not change the routing cost and just add an extra $\epsilon$ to the facility cost. Also, we cannot swap $S$ with $o_i$. If we did this then the facility cost would be just $\epsilon$ but the routing cost would be $3 \cdot (k -1)+1 + (k-1) \cdot (k-1)^d$ since all demand need to be routed at $o_i$. Thus having $S$ as our only facility is a locally optimal solution with a local ratio of $O(k^d)$.

To get more intuition on the difficulty of this approach, following the techniques  in the related literature that are usually employed to provide approximation guarantees,   we would state a list of local steps and try to find out information concerning a locally optimal solution. We will have to use the fact that no local step can improve the cost of the locally optimal state in order to bound its cost. However we need good estimations concerning the new routing costs once a local step is made. Since the edge cost functions depend on the amount of traffic they receive, then we cannot come up with an estimate concerning the routing cost while taking into consideration only demands that are being affected due to the local step. In other words, in order to estimate the increase or decrease in the total routing cost we need to take into account all of the demand that have passed through each edge up until we reached our final state. This fact increases the complexity and new techniques must be created in order to tackle this problem.

\subsection{Why Linear Programming does not Work in our Setting}

A second natural idea would be to find some underlying structure of the optimal solution in order to come up with an LP-rounding or primal-dual algorithm. To do so in this section we will try and simplify our problem. We will assume that the facility cost is the same for all nodes and equal to $B$ (i.e. $f_j = B, \forall j \in V$). We will also assume that our cost functions are linear functions with no negative coefficients (i.e. $l_e(x) = a \cdot x + b$ where $a,b \geq 0, \forall e \in E$).

Consider the following natural formulation of the problem:

\begin{equation*}
    min \sum_{e\in E} (a_{e} \cdot x^2_{e}+b_e \cdot x_e)+\sum_{e\in F}z_e \cdot B
\end{equation*}

\begin{equation*}
    \begin{array}{crcll}
    s.t.&   \sum_{e\in \delta^-(v)}x_{e}&=&  \sum_{e\in \delta^+(v)}x_{e},&   \forall v\in V \\
    &   x_{e}&\leq&  z_e \cdot \sum_{s\in S}w_s,&   \forall e\in F  \\
    & x_{e} &=& w_s & \forall e \in S \\
    &   z_e&\in& \{0,1\}&   \forall e\in F  
\end{array}
\end{equation*}

In the above formulation we treat demands as edges that lead into the graph and carry already $w_s$ demand, and facilities as edges that leave from the graph. The variable $z_e$ becomes $1$ when the corresponding facility is opened and $0$ otherwise. Thus the second inequality of our restrictions ensures that only open facilities are being used. This linear program has an integer restriction and thus, a priori, cannot be solved in polynomial time (unless P=NP). However this restriction is essential to the formulation. Consider the relaxation where we just demand that $0 \leq z_e \leq 1$. This problem can be solved in polynomial time however its optimal solution offer no valuable information because of the following observation.

When relaxing the integrality constraint on $z_e$, since we have a minimization problem $z_e$ will receive the smallest possible value and thus we will have $x_e=z_e\cdot \sum_{s\in S}w_s \Rightarrow z_e=\frac{x_e}{\sum_{s\in S}w_s} \Rightarrow \sum_{e\in F}z_e\cdot B= \sum_{e\in F}\frac{x_e}{\sum_{s\in S}w_s}\cdot B= \frac{B}{\sum_{s\in S}w_s} \cdot \sum_{e\in F} x_e$. However all demands must be served by a facility so $\sum_{e\in F} x_e = \sum_{s\in S}w_s \Rightarrow \frac{B}{\sum_{s\in S}w_s} \cdot \sum_{e\in F} x_e = B$. This tells us that no mater how many facilities are opened and no matter how the demand will be arranged to them, the facility cost will always be equal to $B$. So in order to minimize the routing cost a fractional facility will be opened at every demand bringing the routing cost down to zero. Since the fractional problem returns always the same solution with the same cost, it is of no use for a better approximation than the naive. In other words the intergrality gap of this non-linear program is $O(n)$. Our problem in general becomes really easy when intergrality constraints are removed and thus LP based solutions have big integrality gaps and thus the inherently cannot provide valuable information.

\section{Non-extendability of the Sparsification Technique to the Multi-Source  Setting.}
\label{app:CarathMultiFails}

Suppose 
we have multiple demand points and the cost of opening a facility is common and equals $B$. First of all we cannot try to use sparsification techniques for each demand point separately because this would lead us to an exhaustive search that is exponential in the amount of demand points. However its is obvious that the number of facilities is bounded by the number of demands. So if we could perform exhaustive search exponential in the amount of demands we could simply check all possible facility combination in the first place thus rendering the use of sparsification techniques useless. 

Thus it becomes evident that we have to treat the flow as a whole in order for sparsification techniques to add value to our analysis. However in doing so the results returned by the exhaustive search might not be a feasible solution of the problem. It might also not take into consideration many small demands and close facilities that are crucial to the optimal solution. So once we have used those techniques we then have to find set of facilities that will accommodate large amount of demands which was the exact problem we began with. To better illustrate this argument consider the following example that captures the essence of the difficulty of the problem.

\begin{figure}[h]
\centerline{\includegraphics[scale=.3]{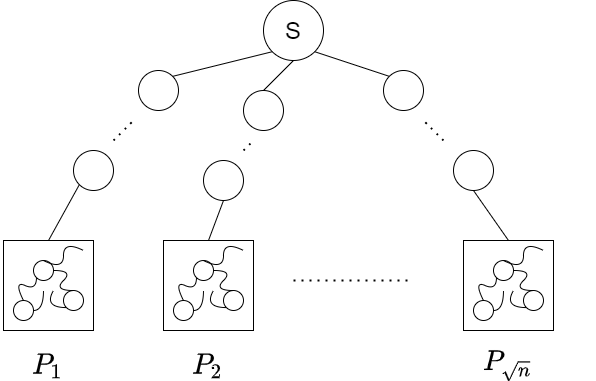}}
\caption{An example where sparsification techniques are rendered useless}
\label{figure:difficultExample}
\end{figure}

If we do not use sparsification techniques the best algorithm we are aware of is the naive $O(n)$-approximate algorithm. Consider the example depicted in figure \ref{figure:difficultExample}. For any $n$ we can create $\sqrt{n}$ different problems where each one of them has $\sqrt{n}$ nodes and $\frac{r}{\sqrt{n}}$ demand (where $r$ is the total demand of the problem). We will call these problems $P_1, P_2, ..., P_{\sqrt{n}}$. Suppose that each problem $P_j$ has $C^*_j$ cost in its optimal solution. We also create a node $S$ and we connect it with each problem with a long chain of nodes. Due to the restrictions of sparsification techniques those chains can be of length at most $b_1 \cdot \log^{b_2}(n)$ where $b_1, b_2$ are constants. This ensures that the optimal solution of the final graph is to solve optimally each problem separately. So the total number of nodes is $\sqrt{n} \cdot \sqrt{n} + \sqrt{n} \cdot b_1 \cdot \log^{b_2}(n) +1 = O(n)$. If we tried to solve each problem separately the best we could do without the use of sparsification techniques would be to use our naive approximate algorithm and end up with a total cost of $\sqrt{n}*OPT$.

Now we will examine how well  sparsification techniques perform in this general problem. Sparsification techniques will examine at most $k$ paths and thus will open at most $k$ facilities. For the exhaustive search of the sparsification techniques to end in quasi-polynomial time we want $k = \textit{poly}(\log(n))$. Thus the sparsification technique will yield a result that addresses at most $k$ from the $\sqrt{n}$ problems. 

If we do not open any more facilities then we will have to route demand from $\sqrt{n}-k$ facilities to $S$ and then to the $k$ addressed problems. This obviously creates a cost that is greater than that of the naive algorithm of opening just one facility in $S$ and routing all demands there. So it becomes apparent that we have to open new facilities. 

The only possible alternative is to solve each of the $\sqrt{n}-k$ unaddressed problems separately using our naive $O(n)$-approximation algorithm. Without loss of generality suppose that $P_1, P_2, ..., P_{\sqrt{n}-k}$ are the unaddressed problems. Solving them naively we get $\sqrt{n} \cdot C^*_1 + \sqrt{n} \cdot C^*_2 + ... + \sqrt{n} \cdot C^*_{\sqrt{n}-k} = \sqrt{n} \cdot \sum_{j=1}^{\sqrt{n}-k}C^*_j$ cost. We will try to find the approximation ratio of our solution. Obviously we can bound from underneath the approximation ratio if we make the following assumptions:

\begin{itemize}
    \item The $k$ addressed problems are solved optimally.
    \item The cost of the $k$ addressed problems is the maximum possible.
    \item The cost of the $\sqrt{n}-k$ unaddressed problems is the minimum possible.
\end{itemize}

We can observe that each problem must open at least one facility and opening one facility in each node is a possible solution (recall that $B$ is the cost of opening a facility). Thus $B \leq C^*_j \leq \sqrt{n} \cdot B$ for all $j$. Thus we get that the approximation ratio of our algorithm is:

\begin{equation*}
    A = \frac{\sqrt{n} \cdot \sum_{j=1}^{\sqrt{n}-k}C^*_j + \sum_{j=\sqrt{n}-k+1}^{\sqrt{n}}C^*_j}{\sum_{j=1}^{\sqrt{n}-k}C^*_j + \sum_{j=\sqrt{n}-k+1}^{\sqrt{n}}C^*_j}
\geq \frac{\sqrt{n} \cdot \sum_{j=1}^{\sqrt{n}-k}B + \sum_{j=\sqrt{n}-k+1}^{\sqrt{n}} \sqrt{n} \cdot B}{\sum_{j=1}^{\sqrt{n}-k}B + \sum_{j=\sqrt{n}-k+1}^{\sqrt{n}}\sqrt{n} \cdot B} 
\end{equation*}

\begin{equation*}
   \Rightarrow   A \geq \frac{\sqrt{n} \cdot (\sqrt{n} - k) + \sqrt{n} \cdot k}{\sqrt{n} - k + \sqrt{n} \cdot k} = \frac{\sqrt{n}}{1- k/\sqrt{n} +k} \geq \frac{\sqrt{n}}{1+k} 
\end{equation*}

Thus $A \geq \frac{\sqrt{n}}{1+k} \geq O(n^{\frac{1}{2}-\lambda})$ for any $\lambda>0$. The second inequality holds since $k = \textit{poly}(\log(n))$. So, especially for large $n$ we have not made any significant progress regarding the $O(\sqrt{n})$ naive algorithm we could have implemented in the first place. Also it is not obvious how we will distinguish those different problems in more difficult settings. Additionally with some hyper parameter tuning such as adjusting the total number of different problems and pose some restraints on the cost of the optimal solution of each problem $P_j$ we can achieve a tighter bound on $A$. Nevertheless this analysis is simple enough to be easily understood but also perfectly illustrates our main point.

\end{document}